\documentclass[letterpaper,conference,10pt]{IEEEtran}
\usepackage[margin=0.625in,right=0.650in,left=0.650in]{geometry}
\makeatletter
\def\ps@headings{%
\def\@oddhead{\mbox{}\scriptsize\rightmark \hfil \thepage}%
\def\@evenhead{\scriptsize\thepage \hfil \leftmark\mbox{}}%
\def\@oddfoot{}%
\def\@evenfoot{}}
\makeatother
 
\usepackage{subcaption}
\usepackage{amssymb,amsmath,amsthm, mathtools}
\usepackage{graphicx}
\usepackage{cite}
\usepackage{latexsym}
\usepackage{verbatim}
\usepackage{amsmath}
\usepackage{amssymb}
\usepackage{caption}
\usepackage{url}
\usepackage{epstopdf}
\usepackage{algorithm}
\usepackage{algcompatible}
\usepackage[noend]{algpseudocode}
\usepackage{times}
\usepackage{stfloats}
\usepackage{multirow}
\usepackage{enumerate}
\usepackage{xspace}
\usepackage{longtable}
\usepackage{mathrsfs}
\usepackage{mathtools}
\usepackage{algcompatible}
\usepackage{fancyhdr}
\usepackage[noend]{algpseudocode}
\usepackage{amssymb}
\usepackage{pifont}
\usepackage{caption}
\usepackage{setspace}
\usepackage{bm}
\usepackage{tikz}

\usepackage{enumitem}

\algrenewcommand\algorithmicindent{.5em}

\newcommand{\pir}{\ensuremath {\mathit{PIR}}{\xspace}}
\newcommand{\itpir}{\ensuremath {\mathit{itPIR}}{\xspace}}
\newcommand{\cpir}{\ensuremath {\mathit{cPIR}}{\xspace}}

\newcommand{\db}{\ensuremath {\mathit{DB}}{\xspace}} 

\newcommand{\crn}{\ensuremath {\mathit{CRN}}{\xspace}}

\newcommand{\su}{\ensuremath {\mathit{SU}}{\xspace}}
\newcommand{\pu}{\ensuremath {\mathit{PU}}{\xspace}}
\newcommand{\dbmatrix}{\ensuremath {\boldsymbol{\mathit{D}}}{\xspace}} 
\newcommand{\dbblock}{\ensuremath {\mathit{b}}{\xspace}} 
\newcommand{\dbrow}{\ensuremath {\mathit{r}}{\xspace}} 
\newcommand{\dbsize}{\ensuremath {\mathit{n}}{\xspace}} 
\newcommand{\dbshare}{\ensuremath {\mathit{\rho}}{\xspace}} 

\newcommand{\ind}{\ensuremath {\mathit{\beta}}{\xspace}}
\mathchardef\mhyphen="2D 

\newcommand{\tp}{\ensuremath {\mathit{t}}{\xspace}} 
\newcommand{\kr}{\ensuremath {\mathit{k}}{\xspace}} 
\newcommand{\wnbr}{\ensuremath {\mathit{s}}{\xspace}} 
\newcommand{\ns}{\ensuremath {\mathit{\ell}}{\xspace}} 
\newcommand{\vbr}{\ensuremath {\mathit{\vartheta}}{\xspace}} 
\newcommand{\bitstr}{\ensuremath {\bm{\mathit{\rho}}}{\xspace}} 
\newcommand{\result}{\ensuremath {\bm{\mathit{R}}}{\xspace}} 
\newcommand{\word}{\ensuremath {\mathit{w}}{\xspace}} 
\newcommand{\x}{\ensuremath {\mathit{l_x}}{\xspace}} 
\newcommand{\y}{\ensuremath {\mathit{l_y}}{\xspace}} 
\newcommand{\chr}{\ensuremath {\mathit{C}}{\xspace}} 
\newcommand{\ts}{\ensuremath {\mathit{ts}}{\xspace}} 
\newcommand{\secret}{\ensuremath {\mathcal{S}}{\xspace}} 
\newcommand{\scode}{\ensuremath {\mathit{S}}{\xspace}} 

\newcommand{\PrSpec}{\ensuremath {\mathit{PriSpectrum}}{\xspace}}
\newcommand{\chorScheme}{\ensuremath {\mathit{LP\mhyphen Chor}}{\xspace}}
\newcommand{\goldbergScheme}{\ensuremath {\mathit{LP\mhyphen Goldberg}}{\xspace}}

\newcommand{\algrule}[1][.2pt]{\par\vskip.5\baselineskip\hrule height #1\par\vskip.5\baselineskip}

\newtheorem{mycorollary}{Corollary}{\bfseries}{\rmfamily}

{\bfseries}{\rmfamily}
\newtheorem{definition}{Definition}{\bfseries}{\rmfamily}
{\bfseries}{\rmfamily}

\IEEEoverridecommandlockouts

\begin{document}


\title{
When the Hammer Meets the Nail: Multi-Server PIR for Database-Driven CRN  with Location Privacy Assurance}


\author{Mohamed Grissa, Attila A. Yavuz, and Bechir Hamdaoui\\
\small Oregon State University, grissam,attila.yavuz,hamdaoui@oregonstate.edu\\ \\ \\ \\
\thanks{This work was supported in part by the US National Science Foundation under NSF award CNS-1162296.

This work has been submitted to the IEEE for possible publication. Copyright may be transferred without notice,
after which this version may no longer be accessible.}
}

\maketitle

\begin{abstract}
We show that it is possible to achieve information theoretic location privacy for secondary users (SUs) in database-driven cognitive radio networks (CRNs) with an end-to-end delay less than a second, which is significantly better than that of the existing alternatives offering only a computational privacy. This is achieved based on a keen observation that, by the requirement of Federal Communications Commission (FCC), all certified spectrum databases synchronize their records. Hence, the same copy of spectrum database is available through multiple (distinct) providers. We harness the synergy between multi-server private information retrieval (PIR) and database-driven CRN architecture to offer an optimal level of privacy with high efficiency by exploiting this observation. We demonstrated, analytically and experimentally with deployments on actual cloud systems that, our adaptations of multi-server PIR outperform that of the (currently) fastest single-server PIR by a magnitude of times with information theoretic security, collusion resiliency and fault-tolerance features. Our analysis indicates that multi-server PIR is an ideal cryptographic tool to provide location privacy in database-driven CRNs, in which the requirement of replicated databases is a natural part of the system architecture, and therefore SUs can enjoy all advantages of multi-server PIR without any additional architectural and deployment costs.
\end{abstract}


\section{Introduction}
\label{sec:Introduction}
The rapid growth of connected wireless devices has dramatically increased the demand for wireless spectrum and led to a serious shortage in spectrum resources. Cognitive radio networks (\crn s)~\cite{mitola1999cognitive} have emerged as a promising technology for solving this shortage problem by enabling dynamic spectrum access (DSA), which improves the spectrum utilization efficiency by allowing unlicensed/secondary users (\su s) to exploit unused spectrum bands (aka spectrum holes or white spaces) of licensed/primary users (\pu s).

Currently, two approaches are being adopted to identify these white spaces: spectrum sensing and geolocation spectrum databases. In the spectrum sensing-based approach, \su s need to sense the \pu~channel to determine whether the channel is available for opportunistic use.
The spectrum database-based approach, on the other hand, does not require that \su s perform sensing to check for spectrum availability. It instead requires that \su s query a database (\db) to learn about spectrum opportunities in their vicinity. This approach, already promoted and adopted by the Federal Communications Commission (FCC), was introduced as a way to overcome the technical hurdles faced by the spectrum sensing-based approaches, thereby enhancing the efficiency of spectrum utilization, improving the accuracy of available spectrum identification, and reducing the complexity
of terminal devices~\cite{gao2013location}.  Moreover, it pushes
the responsibility and complexity of complying with spectrum
policies to \db~and eases the adoption of policy changes
by limiting updates to just a handful number of databases, as
opposed to updating large numbers of devices~\cite{chen2015protocol}.

FCC has designated nine entities (e.g. Google~\cite{google}, iconectiv~\cite{iconectiv}, and Microsoft~\cite{microsoft}) as TV bands device database administrators which are required to follow the guidelines provided by PAWS (Protocol to Access White Space) standard~\cite{chen2015protocol}. PAWS sets guidelines and operational requirements for both the spectrum database and the \su s querying it. These include: \su s need to be equipped with geo-location capabilities, \su s must query \db~with their specific location to check channel availability before starting their transmissions, \db~must register \su s and
manage their access to the spectrum, \db~must respond to \su s' queries with the list of available channels in their vicinity along with the appropriate transmission parameters. As specified by PAWS standard, \su s may be served by several spectrum databases and are required to register to one or more of these databases prior to querying them for spectrum availability. The spectrum databases are reachable via the Internet, and \su s querying these databases are expected to have some form of Internet connectivity\cite{mancuso2013protocol}.

\subsection{Location Privacy Issues in Database-Driven \crn s}

Despite their effectiveness in improving spectrum utilization efficiency, database-driven \crn s suffer from serious security and privacy threats. Since they could be seen as a variant of of {\em location based service (LBS)}, the disclosure of location information of \su s represents the main threat to \su s when it comes to obtaining spectrum availability from \db s. This is simply because {\em \su s have to share their locations with \db s to obtain spectrum availability information in their vicinity.} The fine-grained location, when combined with publicly available information, can easily reveal other personal information about an individual including his/her behavior, health condition, personal habits or even beliefs. For instance, an adversary can learn some information about the health condition of a user by observing that the user regularly goes to a hospital for example. The frequency and duration of these visits can even reveal the seriousness of a user illness and even the type of illness if the location corresponds to that of a specialty clinic. The adversary could even sell this information to pharmaceutical advertisers without the user's consent.

Being aware of such potential privacy threats, \su s may refuse to rely on \db~for spectrum availability information, which may present a serious barrier to the adoption of database-based \crn s, and to the public acceptance and promotion of the dynamic spectrum sharing paradigm. Therefore, {\em there is a critical need for developing techniques to protect the location privacy of \su s while allowing them to harness the benefits of the \crn~paradigm without disrupting the functionalities that these techniques are designed for to promote dynamic spectrum sharing}.


\subsection{Research Gap and Objectives} \label{subsec:ResearchGapObjectives}

Despite the importance of the location privacy issue in \crn s, only recently has it started to gain interest from the research community~\cite{grissa2017location}. Some works focus on addressing this issue in the context of collaborative spectrum sensing~\cite{li2012location,grissa2015location,wangprivacy,grissa2016efficient,grissa2017preserving}; others address it in the context of dynamic spectrum auction~\cite{liu2013location}. 
%
Protecting \su s' location privacy in database-driven \crn s is a more challenging task, merely because \su s are required, by protocol design, to provide their physical location to \db~to learn about spectrum opportunities in their vicinities.
The existing location privacy preservation techniques for database-driven \crn~(e.g.,~\cite{zhang2015optimal,gao2013location,troja2014leveraging,troja2015efficient,zhang2015achieving}) generally rely on three main lines of privacy preserving technologies, (i)   {\em k-anonymity}~\cite{gruteser2003anonymous}, (ii) {\em differential privacy}~\cite{dwork2008differential}  and (iii) single-server {\em Private Information Retrieval (\pir)}~\cite{chor1998private}. However, the direct adaptation of {\em k-anonymity} based techniques have been shown to yield either insecure or extremely costly results~\cite{zang2011anonymization}. The solutions adapting {\em differential privacy} (e.g.,~\cite{zhang2015achieving}) not only incur a non-negligible overhead, but also introduce a noise over the queries, and therefore they may negatively impact the accuracy of spectrum availability information.

Among these alternatives, single-server \pir~seems to be the most popular alternative in the context of \crn s. \pir~technology is a suitable choice for database-driven \crn s, as it permits privacy preserving queries on a public database, and therefore can enable a \su~to retrieve spectrum availability information from the database without leaking his/her location information. However, single-server \pir~protocols rely on highly costly partial homomorphic encryption schemes, which need to be executed over the entire database for each query. Indeed, as we also demonstrated with our experiments in Section \ref{sec:Performance}, the execution of a single query even with some of the most efficient single-server \pir~schemes~\cite{aguilar2016xpir} takes approximately $20$ seconds with a $80\:Mbps/\:30Mbps$ bandwidth on a moderate size database (e.g., $10^6$ entries). An end-to-end delay with the orders of $20$ seconds might be undesirable for spectrum sensing needs of \su s in real-life applications. Also, some of the state-of-the-art efficient computational \pir~schemes~\cite{trostle2010efficient} that are used in the context of \crn s have been shown to be broken~\cite{aguilar2016xpir}. We provide a discussion about the existing privacy enhancing techniques and their potential adaptations to database-driven \crn~settings in Sections~\ref{sec:Performance} and~\ref{sec:Related}.

{\em There is a significant need for practical location privacy preservation approaches for database-driven \crn s that can meet the efficiency and functionality requirements of \su s.}

\subsection{Our Observation and Contribution}
The objective of this paper is to develop efficient techniques for database-driven \crn s that preserve the location privacy of \su s during their process of acquiring spectrum availability information. Specifically, we will aim for the following design objectives: $(i)$ ({\em location privacy}) Preserve the location privacy of \su s while allowing them to receive spectrum availability information; $(ii)$ {\em (efficiency and practicality)} Incur minimum computation, communication and storage overhead. The cryptographic delay must be minimum to permit fast spectrum availability decision for the \su s, and storage/processing cost must be low to enable practical deployments. $(iii)$ {\em (fault-tolerance and robustness)} Mitigate the effects of system failures or misbehaving entities (e.g., colluding databases). {\em It is a very challenging task to meet all of these seemingly conflicting design goals simultaneously.}

The main idea behind our proposed approaches is to harness special properties and characteristics of the database-driven \crn~systems to employ private query techniques that can overcome the significant performance, robustness and privacy limitations of the state-of-the-art techniques. Specifically, our proposed approach is based on the following observation:
%

\textbf{Observation}: {\em FCC requires that all of its certified databases synchronize their records obtained through registration procedures with one another~\cite{fcc2012white,fcc2012TVWS} and need to be consistent across the other databases by providing exactly the same spectrum availability information, in any region, in response to \su s' queries~\cite{ramjee2016critique}. That is, the same copy of spectrum database is available and accessible to the \su s via multiple (distinct) spectrum database administrators/providers.  Is it possible exploit this observation to achieve efficiency location preservation techniques for database-driven \crn?}

In practice,  as stated in PAWS standard~\cite{chen2015protocol}, \su s have the option to register to multiple spectrum databases belonging to multiple service providers.  Currently, many companies (e.g. Google~\cite{google}, iconectiv~\cite{iconectiv}, etc) have obtained authorization from FCC to operate geo-location spectrum databases upon successfully complying to regulatory requirements. Several other companies are still underway to acquire this authorization\cite{fcc-administrators}. Thus, it is more natural and realistic to take this fact into consideration when designing privacy preserving protocols for database-based \crn s. Based on this observation, our main contribution is as follows:
\begin{table}[h!]
\centering  \caption{ Performance Comparison} \label{tab:Table0}

\renewcommand{\arraystretch}{1.5}{
\resizebox{.47\textwidth}{!}{%
\begin{tabular}{||c||c||c|c|c||c||}

\hline {\multirow{2}{*}{\textbf{Scheme} }}   & {\multirow{2}{*}{\textbf{Comm.}}} & \multicolumn{3}{|c||}{\textbf{Delay}} & {\multirow{2}{*}{\textbf{Privacy}}}\\ \cline{3-5}
 &  &  {$\boldsymbol \db$} & {$\boldsymbol \su$}& \textbf{ end-to-end}& \\ \hline

\hline \hline  \chorScheme &  $ 753\:KB $ &  $0.48\:s$ & $0.008\:s$ & $0.62\:s$& $(\ns-1)${\em-private}\\

\hline  \goldbergScheme & $ 6000\:KB$ & $1.21\:s $ & $ 0.32\: s$ & $1.78\:s$& \tp{\em-private \ns-comp.-private} \\

\hline  \hline \PrSpec~\cite{gao2013location} & $ 512.8\: KB $  & $21\:s $ & $0.084 \:s$& $24.2$ & {\em underlying \pir~ broken}\\

\hline  Troja et al~\cite{troja2015efficient} & $ 8.4\:KB$ & $11760\:s$ & $5.62\:s$ & $11766\:s$ & {\em computationally-private}\\

\hline  Troja et al~\cite{troja2014leveraging} & $12120\:KB $ & $11760\:s$ & $48\:s$& $11820\:s$ & {\em computationally-private}\\

\hline \hline XPIR~\cite{aguilar2016xpir} & $ 4321\:KB $ & $17.66\:s$ & $ 0.34\:s$ & $20.53\:s$ & {\em computationally-private}\\ \hline
\end{tabular}}
}

\flushleft{\scriptsize{\textbf{Parameters:} $\dbsize = 560\:MB,\:\dbblock = 560\:B, \:\dbrow = 10^6, \:\ns=6, \:\word = 8, \;\kr = 6$
}}
\end{table}


\textbf{Our Contribution}: {\em To the best of our knowledge, we are the first to exploit the observation that multiple copies of spectrum \db s are available by nature in database-driven \crn s, and therefore it is possible to harness  multi-server \pir~techniques~\cite{chor1998private,goldberg2007improving} that offer information-theoretic privacy with substantial efficiency advantages over single-server \pir. We show, analytically and experimentally with deployments on cloud systems, that our adaptation of multi-server \pir~techniques significantly outperforms the state-of-the-art location privacy preservation methods as demonstrated in Table~\ref{tab:Table0} and detailed in Section \ref{sec:Performance}. Moreover, our adaptations achieve the information theoretical privacy while existing alternatives offer only computational privacy. This feature provides an assurance against even post-quantum adversaries~\cite{chen2016nistir} and can avoid recent attacks on computational \pir~\cite{aguilar2016xpir}.}

Notice that, multi-server \pir~techniques require the availability of multiple (synchronized) replicas of the database. Therefore, despite their high efficiency and security, they received a little attention from the practitioners. For instance, in traditional data outsourcing settings (e.g., private cloud storage), the application requires a client to outsource only a single copy of its database. The distribution and maintenance of multiple copies of the database across different service providers brings additional architectural and deployment costs, which might not be economically attractive for the client.

In this paper, we showcased one of the first natural use-cases of multi-server \pir, in which the multiple copies of synchronized databases are already available by the original design of application (i.e., spectrum availability information in multi-database \crn s), and therefore multi-server \pir~does not introduce any extra overhead on top of the application. Exploiting this synergy between multi-database \crn~and multi-server \pir~permitted us to provide informational theoretical location privacy for \su s with a significantly better efficiency compared to existing single-server \pir~approaches.

\textbf{Desirable Properties:} We outline the desirable properties of our approaches below.

\begin{itemize}[leftmargin=*]
\item \noindent{\em Computational efficiency:} The adapted approaches are much more efficient than existing location privacy preserving schemes. For instance, as shown in Table~\ref{tab:Table0}, \chorScheme~and \goldbergScheme~are more than $3$ orders of magnitudes faster than the schemes proposed by Troja et al.~\cite{troja2014leveraging,troja2015efficient}, and $10$ times faster than XPIR~\cite{aguilar2016xpir} and \PrSpec~\cite{gao2013location}.
\item {\em Information Theoretical Privacy Guarantees:} They can achieve information-theoretic privacy which is the optimal privacy level that could be reached as opposed to computational privacy guarantees offered by existing approaches. In fact some of these approaches are prone to recent attacks on computational-\pir~protocols~\cite{aguilar2016xpir} and are not secure against post-quantum adversaries~\cite{chen2016nistir}.
\item {\em Low communication overhead:} Both approaches provide a reasonable communication overhead that is a middle ground between the fastest computational \pir~\cite{aguilar2016xpir} and the most communication efficient computational \pir~\cite{gentry2005single}.
    \item {\em Fault-Tolerance and Robustness}: Our proposed approaches are resilient to the issues that are associated with multi-server architectures: failures, byzantine behavior, and collusion. Both \chorScheme~and \goldbergScheme~can handle collusion of multiple \db s. In addition, \goldbergScheme~can also handle faulty and byzantine~\db s.
 \item {\em Experimental evaluation on actual cloud platforms}: We deploy our proposed approaches on a real cloud platform, GENI~\cite{Berman20145}, to show their feasibility. In our experiment, we create multiple geographically distributed VMs each playing the role of a \db. A laptop plays the role of a \su~that queries \db s, i.e. VM s. Our experiments confirm the superior computational advantages of the adaption of multi-server \pir~over the existing alternatives.
\end{itemize}

\section{Preliminaries and Models}
\label{sec:Prelim}
\subsection{Notation and Building Blocks}
We summarize our notations in Table~\ref{t:notations}. Our adaptations of multi-server \pir~rely on the following building blocks.

\begin{table}[h!]
\caption{\small Notations}
\centering
\resizebox{0.4\textwidth}{!}{
\label{t:notations}
\begin{tabular}{l l}

\hline
\noalign{\medskip }
$\db$ & Spectrum database \\
$\su$ & Secondary user \\
$\crn$ & Cognitive radio network \\
$\ns$ & Number of spectrum databases \\
$\dbmatrix$ & Matrix modeling the content of \db \\
$\dbrow$ & Number of records in \dbmatrix \\
$\dbsize$ & Size of the database in bits\\
$\dbblock$ & Size of one record of the database in bits\\
$\word$ & Size of one word of the database in bits\\
$\wnbr$ & Number of words per block\\
$\ind$ & Index of the record sought by \su\\
$\tp$ & Privacy level (tolerated number of colluding \db s)\\
$\kr$ & Number of responding \db s\\
$\vbr$ & Number of byzantine \db s\\
\noalign{\smallskip} \hline \noalign{\smallskip}
\end{tabular}
}
\end{table}

\noindent {\bf Private Information Retrieval ($\boldsymbol \pir$):} \pir~allows a user to retrieve a data item of its choice from a database, while preventing the server owning the database from gaining information on the identity of the item being retrieved~\cite{beimel2001information}. One trivial solution to this problem is to make the server send an entire copy of the database to the querying user. Obviously, this is a very inefficient solution to the \pir~problem as its communication complexity may be prohibitively large. However, it is considered as the only protocol that can provide information-theoretic privacy, i.e. perfect privacy, to the user's query in single-server setting. There are two main classes of \pir~protocols according to their privacy level: information-theoretic \pir~(\itpir) and computational \pir~(\cpir).
\begin{itemize}[leftmargin=*]
\item \noindent {\em Information-theoretic or multi-server \pir:} It guarantees information-theoretic privacy to the user, i.e. privacy against computationally unbounded servers. This could be achieved efficiently only if the database is replicated at $k \geq 2$ non-communicating servers~\cite{chor1998private,goldberg2007improving}. The main idea behind these protocols consists on decomposing each user's query into several sub-queries to prevent leaking any information about the user's intent.
\item \noindent {\em Computational or single-server \pir:} It guarantees privacy against computationally bounded server(s). In other words, a server cannot get any information about the identity of the item retrieved by the user unless it solves a certain computationally hard problem (e.g. factoring a large prime), which is common in modern cryptography. Thus, they offer weaker privacy than their \itpir~counterparts~\cite{trostle2010efficient,melchor2008fast}.
\end{itemize}
\noindent {\bf Shamir Secret Sharing:} This is a concept introduced by Shamir et al.~\cite{shamir1979share} to allow a secret holder to divide its secret \secret~into \ns~shares $\secret_1, \cdots, \secret_{\ns}$ and distribute these shares to \ns~parties. In $(\tp,\ns)$-Shamir secret sharing, where $\tp < \ns$, if \tp~or fewer combine their shares, they learn no information about \secret. However, if more than \tp~come together, they can easily recover \secret. Given a secret \secret~chosen arbitrarily form a finite field, the $(\tp,\ns)$-Shamir secret sharing scheme works as follows: the secret holder chooses \ns~arbitrary non-zero distinct elements $\alpha_1,\cdots,\alpha_{\ns} \in \mathbb{F}$. Then, it selects \tp~elements $\sigma_1,\cdots,\sigma_{\tp}\in \mathbb{F}$ uniformly at random. Finally, the secret holder constructs the polynomial $f(x) = \sigma_0 + \sigma_1 x + \sigma_2 x^2 + \cdots + \sigma_t x^t$, where $\sigma_0 = \secret$. The \ns~shares $\secret_1, \cdots, \secret_{\ns}$, that are given to each party, are $(\alpha_1,f(\alpha_1)),\cdots,(\alpha_{\ns},f(\alpha_{\ns}))$. Any $\tp+1$ or more parties can recover the polynomial $f$ using Lagrange interpolation and thus they can reconstruct the secret $\secret = f(0)$. However, \tp~or less parties can learn nothing about \secret. In other words, if $\tp+1$ shares of \secret~are available then \secret~can be easily recovered.

\subsection{System Model and Security Definitions}
We consider a database-driven \crn~that contains \ns~\db s, where $\ns\geq 2$, and a \su~registered to these \db s to learn spectrum availability information in its vicinity. We assume that these \db s share the same content and that they are synchronized as mandated by PAWS standard~\cite{chen2015protocol}. We also assume that \db s may collude in order to infer \su's location. In the following, we present our security definitions.
\begin{definition}\label{def:byzantine}
\textbf{\em Byzantine} {\boldsymbol \db}{\bf:} This is a faulty \db~that runs but produces incorrect answers, possibly chosen maliciously or computed in error. This might be due to a corrupted or obsolete copy of the database caused by a synchronization problem with the other \db s.
\end{definition}

\begin{definition}\label{def:t-private}
\textbf{\tp{\em -private}} {\boldsymbol\pir}{\bf:} The privacy of the query is information-theoretically protected, even if up to \tp~of the \ns~\db s collude, where $\tp < \ns$.
\end{definition}

\begin{definition}\label{def:v-byzantine-robustness}
\textbf{\vbr{\em -Byzantine-robust}} {\boldsymbol\pir}{\bf:} Even if \vbr~of the responding \db s are {\em Byzantine}, \su~can reconstruct the correct database item, and determine which of the \db s provided incorrect response.
\end{definition}

\begin{definition}\label{def:k-robustness}
\textbf{{\em \kr-out-of-\ns}} {\boldsymbol\pir}{\bf:} \su~can reconstruct the correct record if it receives at least \kr-out-of-\ns~responses, $2\leq\kr\leq\ns$.
\end{definition}

\begin{definition}\label{def:robustness}
\textbf{{\em Robust}} {\boldsymbol\pir}{\bf:} It can deal with \db s that do not respond to \su's queries and allows \su~to reconstruct the correct output of the queries in this situation.
\end{definition}

\section{Proposed Approaches}
In the proposed approaches, we tailor multi-server \pir~to the context of multi-\db~\crn s. We start by illustrating the structure of the spectrum database that we consider. Then, we give two approaches, each adapts a multi-server \pir~protocol with different security and performance properties.
We model the content of each \db~as an $\dbrow \times \wnbr$ matrix \dbmatrix~of size $\dbsize$ bits, where $\wnbr$ is the number of words of size \word~in each record/block of the database and $\dbrow$ is the number of records in the database, i.e. $\dbrow = \dbsize/\dbblock$, where $\dbblock=\wnbr\times\word$ is the block size in bits. The $k^{th}$ row of \dbmatrix~is the $k^{th}$ record of the database. 
\[ \dbmatrix = 
\begin{bmatrix}
    \word_{11} & \word_{12}  & \dots  & \word_{1\wnbr} \\
    \word_{21} & \word_{22}  & \dots  & \word_{2\wnbr} \\
    \vdots & \vdots  & \ddots & \vdots \\
    \word_{\dbrow 1} & \word_{d2}  & \dots  & \word_{\dbrow\wnbr}
\end{bmatrix}
\]
We further assume that each row of the database corresponds to a unique combination of the tuple $(\x,\y,\chr,\ts)$, where \x~and \y~represent one location's latitude and longitude, respectively, \chr~is a channel number, and \ts~is a time-stamp. We also assume that \su s can associate their location information with the index \ind~of the corresponding record of interest in the database using some inverted index technique that is agreed upond with \db s. An \su~that wishes to retrieve record $\dbmatrix_\ind$ without any privacy consideration can simply send to \db~a row vector $e_\ind$ consisting of all zeros except at position \ind~where it has the value $1$. Upon receiving $e_\ind$, \db~multiplies it with \dbmatrix~and sends record $\dbmatrix_\ind$ back to \su~as we illustrate below: 
\[\begin{bmatrix}
    0 & \dots  & 0  & 1 & 0 & \dots & 0 \\
\end{bmatrix} 
\begin{bmatrix}
    \word_{11} & \word_{12}  & \dots  & \word_{1\wnbr} \\
    \word_{21} & \word_{22}  & \dots  & \word_{2\wnbr} \\
    \vdots & \vdots  & \ddots & \vdots \\
    \word_{\dbrow 1} & \word_{d2}  & \dots  & \word_{\dbrow\wnbr}
\end{bmatrix}\]
\[= 
\begin{bmatrix}
    \word_{\ind 1} & \word_{\ind 2}  & \dots  & \word_{\ind\wnbr} \\
\end{bmatrix}\]
This trivial approach makes it easy for \db s to learn \su's location from the vector $e_\ind$ as \dbmatrix~is indexed based on location. In the following we present two approaches that try to hide the content of $e_\ind$ from \db s, and thus preserve \su's location privacy. The approaches present a tradeoff between efficiency, and some additional security features.

\subsection{Location Privacy with Chor (\chorScheme)}

\begin{figure}[h!]
\centering
\includegraphics[scale=0.3]{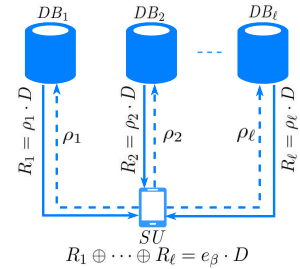} 
\caption{Main steps of \chorScheme~Algorithm}
\label{fig:chor}
\end{figure}

Our first approach, termed \chorScheme, harnesses the simple and efficient \itpir~protocol proposed by Chor et al.~\cite{chor1998private}. We describe the different steps of \chorScheme~in Algorithm~\ref{alg:chor} and highlight these steps in \figurename~\ref{fig:chor}. Elements of \dbmatrix~in this scheme belong to $GF(2)$, i.e. $\word = 1$ bit and $\dbblock = \wnbr$. 

\begin{algorithm}[h!]
\caption{$\dbmatrix_\ind\gets$\chorScheme(\ns, \dbrow, \dbblock)}\label{alg:chor}
\begin{algorithmic}[1]

\Statex \pmb \su 
\State $\ind \gets InvIndex(\x,\y,\chr,\ts)$
\State Sets standard basis vector $\bm{e}_\ind \gets \overrightarrow{1}_\ind \in \mathbb{Z}^\dbrow$
\State Generates $\bitstr_1, \cdots, \bitstr_{\ns-1} \in_R GF(2)^\dbrow$
\State $\bitstr_{\ns} \gets \bitstr_1 \oplus \cdots \oplus \bm{e}_\ind$
\State Sends $\bitstr_i$ to $\db_i$, for $1 \leq i \leq \ns$
\hspace{20pt}\algrule
\Statex {\bf Each $\pmb\db_i$}
\State Receives $\bitstr_i = \bitstr_{i1}\cdots\bitstr_{i\dbrow} \in \{0,1\}^\dbrow$
\State $\result \gets \bigoplus\limits_{\substack{1\leq j\leq \dbrow \\ \bitstr_{ij}=1}}\dbmatrix_j$, $\dbmatrix_j$ is the $j^{th}$ block of \dbmatrix
\State Sends $\result_i$ to \su
\hspace{20pt}\algrule
\Statex \pmb \su
\State Receives $\result_1,\cdots,\result_{\ns}$
\State $\dbmatrix_{\ind} \gets \result_1 \oplus\cdots\oplus \result_{\ns}$
\end{algorithmic}
\end{algorithm}
 In \chorScheme, \su~starts by invoking the inverted index subroutine $InvIndex(\x,\y,\chr,\ts)$ which takes as input the coordinates of the user, its channel of interest, and a time-stamp and returns a value \ind. This value corresponds to the index of the record $\dbmatrix_\ind$ of \dbmatrix~that \su~is interested in. \su~then constructs $\bm{e}_\ind$, which is a standard basis vector $\overrightarrow{1}_\ind \in \mathbb{Z}^\dbrow$ having $0$ everywhere except at position $\ind$ which has the value $1$ as we discussed previously. \su~also picks $\ns-1$ \dbrow-bit binary strings $\bitstr_1, \cdots, \bitstr_{\ns-1}$ uniformly at random from $GF(2)^\dbrow$, and computes $\bitstr_{\ns} = \bitstr_1 \oplus \cdots \oplus \bm{e}_\ind$. Finally, \su~sends $\bitstr_i$ to $\db_i$, for $1 \leq i \leq \ns$. Upon receiving the bit-string $\bitstr_i = \bitstr_{i1} \oplus \cdots \bitstr_{i\dbrow}$ of length $\dbrow$, $\db_i$ computes $\result_i = \bitstr_i \cdot \dbmatrix$, which could be seen also as the XOR of those blocks $\dbmatrix_j$ in $\dbmatrix$ for which the $j^{th}$ bit of $\bitstr_i$ is $1$, then sends $\result_i$ back to \su. \su~receives $\result_i$s from $\db_i$s, $1 \leq i \leq \ns$, and computes $\result_1 \oplus \cdots \oplus \result_{\ns} = (\bitstr_1 \oplus \cdots \oplus \bitstr_{\ns}) \cdot \dbmatrix = \bm{e}_\ind \cdot \dbmatrix$, which is the $\ind^{th}$ block of the database that \su~is interested in, from which it can retrieve the spectrum availability information. 

\chorScheme~is very efficient thanks to its reliance on simple XOR operations only as we discuss in Section~\ref{sec:Performance}. It is also {$(\ns-1)$\em-private}, by Definition~\ref{def:t-private}, as collusion of up to $\ns-1$ \db s cannot enable them to learn $\bm{e}_\ind$, and consequently its location. In fact, only if \ns~\db s collude, then they will be able to learn $\bm{e}_\ind$ by simply XORing their $\{\bitstr_i\}_{i=1}^{\ns}$. However this approach suffers from two main drawbacks. First, it is not {\em robust} since even if one \db~fails to respond, \su~will not be able to recover $\dbmatrix_\ind$. Second, it is not {\em byzantine robust}; if one or more \db s return a wrong response, \su~will reconstruct a wrong block and also will not be able to recognize which \db~misbehaved so as not to rely on it for future queries. In Section~\ref{sec:goldberg} we discuss a second approach that improves on these two aspects but with some additional overhead.

\subsection{Location Privacy with Goldberg (\goldbergScheme)}\label{sec:goldberg}
\begin{figure}[h!]
\center
\includegraphics[scale=0.34]{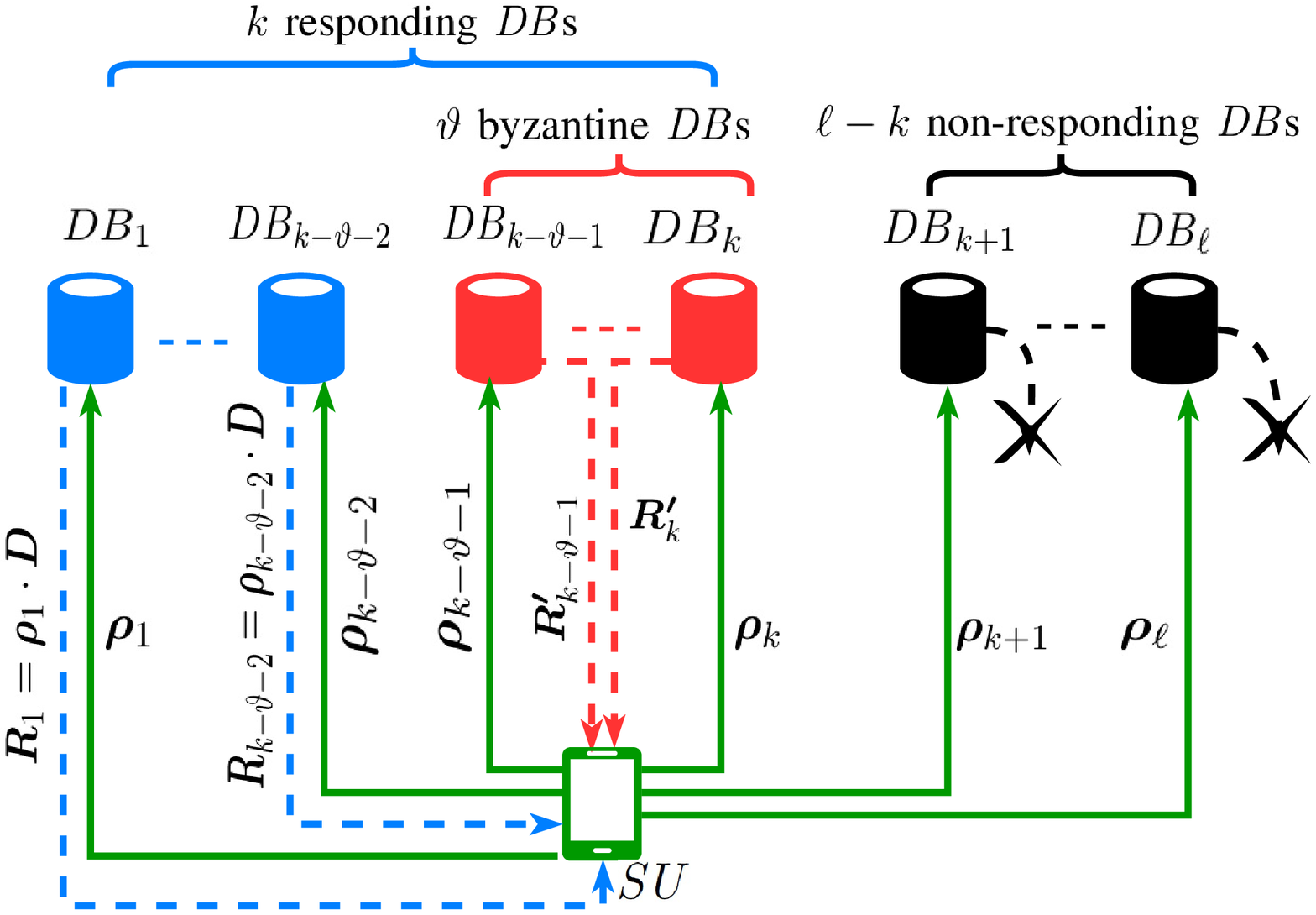} 
\caption{Illustration of \goldbergScheme}
\label{fig:pirGoldberg}
\end{figure}

Our second approach, termed \goldbergScheme, is based on Goldberg's \itpir~protocol~\cite{goldberg2007improving} which uses Shamir secret sharing to hide $\bm{e}_\ind$, i.e. \su's query. It is a modification of Chor's scheme~\cite{chor1998private} to achieve both {\em robustness} and {\em byzantine robustness}. Rather than working over $GF(2)$ (binary arithmetic), this scheme works over a larger field $\mathbb{F}$, where each element can represent $w$ bits. The database $\dbmatrix = (\word_{jk}) \in \mathbb{F}^{\dbrow \times \wnbr}$ in this scheme, is an $\dbrow \times \wnbr$ matrix of elements of $\mathbb{F} = GF(2^w)$. Each row represents one block of size $\dbblock$ bits, consisting of $\wnbr$ words of $\word$ bits each. Again, $\dbmatrix$ is replicated among $\ns$ databases $\db_i$. We summarize the main steps of \goldbergScheme~protocol in Algorithm~\ref{alg:goldberg} and illustrate them in \figurename~\ref{fig:pirGoldberg}.
  
To determine the index \ind~of the record that corresponds to its location, \su~starts by invoking the subroutine  $InvIndex(\x,\y,\chr,\ts)$ then constructs the standard basis vector $\bm{e}_\ind \in \mathbb{F}^r$ as explained earlier. \su~then uses $(\ns,\tp)$-Shamir secret sharing to divide the vector $\bm{e}_\ind$ into $\ns$ independent shares $(\alpha_1,,\bm{\dbshare}_1)\cdots,(\alpha_\ns,\bm{\dbshare}_\ns)$ to ensure a $\tp${\em-private} \pir~protocol as in Definition~\ref{def:t-private}. That is, \su~chooses $\ns$ distinct non-zero elements $\alpha_i \in \mathbb{F}^*$ and creates $\dbrow$ random degree-$\tp$ polynomials $f_1,\cdots, f_\dbrow$ satisfying $f_j(0) = \bm{e}_\beta[j]$. \su~then sends to each $\db_i$ its share corresponding to the vector $\bm{\dbshare}_i = \langle f_1(\alpha_i),\cdots, f_r(\alpha_i)\rangle$. Each $\db_i$ then computes the product $\bm{\result}_i = \bm{\dbshare}_i \cdot \dbmatrix = \langle\sum_j f_j(\alpha_i)\bm{\word}_{j1}, \cdots, \sum_j f_j(\alpha_i)\bm{\word}_{js}\rangle \in \mathbb{F}^s$ and sends $\bm{\result}_i$ to \su.

Some \db s may fail to respond to \su's query and only \kr-out-of-\ns~send their responses to \su. \su~collects \kr~responses from the \kr~responding \db s and tries to recover the record at index $\beta$ from the $\result_i$s by using the $EasyRecover()$ subroutine from~\cite{goldberg2007improving} which uses Lagrange interpolation to recover $\dbmatrix_\ind$ from the secret shares $(\alpha_1,\result_1),\cdots,(\alpha_\kr,\result_{\kr})$. This is possible thanks to the use of $(\ns,\tp)$-Shamir secret sharing as long as $\kr > \tp$ and these \kr~\db s are honest. In fact, by the linearity property of Shamir secret sharing, since $\{(\alpha_i,\bm{\dbshare}_i)\}_{i=1}^{\ell}$ is a set of $(\ns,\tp)$-Shamir secret shares of $\bm{e}_\beta$, then $\{(\alpha_i, \result_i)\}_{i=1}^{\ell}$ will be also a set of $(\ns,\tp)$-Shamir secret shares of $\bm{e}_\beta \cdot \dbmatrix$, which is the $\beta^{th}$ block of the database. Thus, it is possible for \su~to reconstruct $\dbmatrix_\ind$ using Lagrange interpolation as explained in Section~\ref{sec:Prelim}, by relying only on the \kr~responses which makes \goldbergScheme~robust by Definition~\ref{def:robustness}. Also, the $EasyRecover()$ can detect the \db s that responded honestly, thus those that are byzantine as well, which should discourage \db s from misbehaving. More details about this subroutine could be found in~\cite{goldberg2007improving}.
\begin{algorithm}[h!]
\caption{$\dbmatrix_\ind\gets\goldbergScheme(\ns, \dbrow, \dbblock, \tp, \word)$}\label{alg:goldberg}

\begin{algorithmic}[1]
\Statex \pmb \su
\State $\ind \gets InvIndex(\x,\y,\chr,\ts)$
\State Sets standard basis vector $\bm{e}_\ind \gets \overrightarrow{1}_\ind \in \mathbb{Z}^\dbrow$
\State Chooses $\ell$ distinct $\alpha_1,\cdots,\alpha_\ns \in \mathbb{F}^*$
\State Creates $r$ random degree-$t$ polynomials $f_1,\cdots, f_r \in_R \mathbb{F}[x]$ s.t. $f_j(0) = \bm{e}_\beta[j]$' $\forall j \in [1,\cdots,\dbrow]$
\State $\bm{\dbshare}_i \gets \langle f_1(\alpha_i),\cdots, f_r(\alpha_i)\rangle$, $\forall i \in [1,\cdots,\ns]$
\State Sends $\bm{\dbshare}_i$ to $\db_i$, $\forall i \in [1,\cdots,\ns]$
\hspace{20pt}\algrule
\Statex {\bf Each honest $\pmb\db_i$}
\State Receives $\bm{\dbshare}_i$
\State $\bm{\result}_i \gets \bm{\rho}_i \cdot \bm{D} = \langle\sum_j f_j(\alpha_i)\bm{w}_{j1}, \cdots, \sum_j f_j(\alpha_i)\bm{w}_{js}\rangle$
\State Sends $\bm{\result}_i$ to \su
\hspace{20pt}\algrule
\Statex \pmb \su
\State Receives $\result_1,\cdots,\result_{\kr}$
\If{$\kr >\tp$}
\State $\dbmatrix_\ind \gets EasyRecover(\tp,\word,[\alpha_1,...,\alpha_\kr],[\result_1,\cdots,\result_{\kr}])$
\ElsIf{Recovery fails \textbf{and} $\vbr< \kr - \lfloor\sqrt{\kr \tp}\rfloor$}
\State $\scode_q \gets \langle \result_1[q],\cdots,\result_\kr[q]\rangle$, $\forall q \in [1,\wnbr]$
\State $\dbmatrix_\ind\gets HardRecover(\tp,\word,[\alpha_1,...,\alpha_\kr],[\scode_1,\cdots,\scode_\wnbr])$
\EndIf

\end{algorithmic}
\end{algorithm}

Moreover, $\vbr$ \db s among the \kr~responding ones may even be {\em byzantine}, as in Definition~\ref{def:byzantine}, and  produce incorrect response. In that case, it would be impossible for \su~to simply rely on Lagrange interpolation to recover the correct responses. Since Shamir secret sharing is based on polynomial interpolation, the problem of recovering the response in the case of {\em byzantine} failures corresponds to noisy polynomial reconstruction, which is exactly the problem of decoding Reed-Solomon codes~\cite{devet2012optimally}. Thus, \su~would rather rely on error correction codes and more precisely on the Guruswami-Sudan list decoding~\cite{guruswami1998improved} algorithm which can correct $\vbr < \kr - \lfloor\sqrt{\kr \tp}\rfloor$ incorrect responses. In fact, the vector $\langle \bm{\result}_1[q], \bm{\result}_2[q],\cdots, \bm{\result}_\ell[q]\rangle$ is a Reed-Solomon code-word encoding the polynomial $g_q = \sum_j f_j \bm{\word}_{jq}$, and the client wishes to compute $g_q(0)$ for each $1 \leq q \leq \wnbr$ to recover all the \wnbr~words forming the record $\dbmatrix_\ind = \langle g_1(0),\cdots,g_\wnbr(0)\rangle$. This is done through the $HardRecover()$ subroutine from~\cite{goldberg2007improving}. This makes \goldbergScheme~also \vbr{\em -Byzantine-robust}, by Definition~\ref{def:v-byzantine-robustness}, and solves the robustness issues that \chorScheme~suffers from, however, this comes at the cost of an additional overhead as we discuss in Section~\ref{sec:Performance}.

\begin{mycorollary}
 \chorScheme~and \goldbergScheme~directly inherit the security properties of Chor's~\cite{chor1998private} \pir~and Goldberg's~\cite{goldberg2007improving} \pir~respectively.
\end{mycorollary}

\section{Evaluation and Analysis}
\label{sec:Performance}
\subsection{Analytical Comparison }

We start by studying \chorScheme~and \goldbergScheme's performance analytically and we compare them to existing approaches. For \goldbergScheme, we choose $\word = 8$ to simplify the cost of computations as in~\cite{devet2012optimally}; since in $GF(2^8)$, additions are XOR operations on bytes and multiplications are lookup operations into a $64$ KB table~\cite{devet2012optimally}. We summarize the system communication complexity and the computation incurred by both \db~and \su~and we illustrate the difference in architecture and privacy level of the different approaches in Table~\ref{tab:Table1}. As we mentioned earlier, existing research focuses on the single \db~setting. We compare \chorScheme~and \goldbergScheme~to these approaches despite the difference of architecture to show the great benefits that multi-server \pir~brings in terms of performance and privacy as we discuss next. We briefly discuss these approaches in the following. 

Gao et al.~\cite{gao2013location} propose a \pir-based approach, termed \PrSpec, that relies on the \pir~scheme of Trostle et al.~\cite{trostle2010efficient} to defend against the new attack that they identify. This new attack exploits spectrum utilization pattern to localize \su s. Troja et al.~\cite{troja2014leveraging,troja2015efficient} propose two other \pir-based approaches that try to minimize the number of \pir~queries by either allowing \su s to share their availability information with other \su s~\cite{troja2014leveraging} or by exploiting trajectory information to make \su s retrieve information for their current and future positions in the same query~\cite{troja2015efficient}.

Despite their merit in providing location privacy to \su s these \pir-based approaches incur high overhead especially in terms of computation. This is due to the fact that they rely on \cpir~protocols to provide location privacy to \su s, which are known to suffer from expensive computational cost. In fact, answering an \su's query through a \cpir~protocol, requires \db~to process all of its records, otherwise \db~would learn that \su~is not interested in them and would then learn partial information about the record~$\dbmatrix_\ind$, and consequently \su's location. This makes the computational cost of most \cpir~based location preserving schemes linear on the database size from \db~side as we illustrate in Table~\ref{tab:Table1}. Now this is not exclusive to \cpir~protocols as even \itpir~protocols may require processing all the records to guarantee privacy, however, the main difference with \cpir~protocols is that the latter have a very large cost per bit in the database, usually involving expensive group operations like multiplication over a large modulus~\cite{aguilar2016xpir} as opposed to multi-server \itpir~protocols. This could be seen clearly in Table~\ref{tab:Table1} as both \chorScheme~and \goldbergScheme~require \db~to perform a very efficient XOR operation per bit of the database. The same applies to the overhead incurred by \su~which only performs XOR operations in both \chorScheme~and \goldbergScheme, while performing expensive modular multiplications and even exponentiations over large primes in the \cpir-based approaches. 

In terms of communication overhead, the proposed approaches incur a cost that is linear in the number of records $\dbrow$ and their size $\dbblock$. As an optimal choice of these parameters is usually $\dbrow = \dbblock = \sqrt{\dbsize}$~\cite{chor1998private,goldberg2007improving,devet2012optimally,aguilar2016xpir} then this cost could be seen as $\mathcal{O}(\sqrt{\dbsize\word})$ to retrieve a record of size $\sqrt{\dbsize\word}$ bits, which is a reasonable cost for an information theoretic privacy.

Moreover, as illustrated in Table~\ref{tab:Table1}, existent approaches fail to provide information theoretic privacy as the underlying security relies on computational \pir~schemes. The only approaches that provide information theoretic location privacy are \chorScheme~and \goldbergScheme~which are $(\ns-1)${\em -private} and {\em\tp-private}, respectively, by Definition~\ref{def:t-private}. It is worth mentioning that \PrSpec~\cite{gao2013location} relies on the well-known \cpir~of Trostle et al.~\cite{trostle2010efficient} representing the state-of-the-art in efficient \cpir. However, this \cpir~scheme has been broken~\cite{aguilar2016xpir,lepoint2015cryptanalysis}. Since the security of \PrSpec~follows that of Trostle et al.~\cite{trostle2010efficient} broken \cpir, then \PrSpec~fails to provide the privacy objective that it was designed for. However, we include it in our performance analysis for completeness. 

\subsection{Experimental Evaluation}
We further evaluate the performance of the proposed schemes experimentally to confirm the analytical observations.

\noindent \textbf{Hardware setting and configuration.} We have deployed the proposed approaches on GENI~\cite{Berman20145} cloud platform using the percy++ library~\cite{percy}. We have created $6$ virtual machines (VMs), each playing the role of a \db~and they all share the same copy of \dbmatrix. We deploy these GENI VMs in different locations in the US to count for the network delay and make our experiment closer to the real case scenario where spectrum service providers are located in different locations. These VMs are running Ubuntu $14.04$, each having $8$ GB of RAM, $15$ GB SSD, and $4$ vCPUs, Intel Xeon X5650 \@~$2.67$ GHz or Intel Xeon E5-2450\@~$2.10$ GHz. To assess the \su~overhead we use a Lenovo Yoga 3 Pro laptop with $8$ GB RAM running Ubuntu $16.10$ with an Intel Core m Processor 5Y70 CPU\@~$1.10$ GHz. The client laptop communicates with the remote VMs through ssh tunnels. We are also aware of the advances in \cpir~technology, and more precisely the fastest \cpir~protocol in the literature which is proposed by Aguilar et al.\cite{aguilar2016xpir}. We include this protocol in our experiment to illustrate how multi-server \pir~performs against the best known \cpir~scheme if it is to be deployed in \crn s. We use the available implementation of this protocol provided in~\cite{Xpir} and we deploy its server component on a remote GENI VM while the client component is deployed on the Lenovo Yoga 3 Pro laptop.

\noindent \textbf{Dataset.} Spectrum service providers (e.g. Google, Microsoft, etc) offer only graphical web interfaces to their databases that return basic spectrum availability information for a user-specified location. Access to real data from real spectrum databases was not possible, thus, we generate random data for our experiment. The generated data consists of a matrix that models the content of the database, \dbmatrix, with a fixed block size $\dbblock = 560$ kB while varying the number of records $\dbrow$. The value of $\dbblock$ is estimated based on the public raw data provided by FCC~\cite{cdbs} on a daily basis and which service providers use to populate their spectrum databases. 

\begin{table*}[ht!]

\centering  \caption{ Comparison with existent schemes} \label{tab:Table1}

\renewcommand{\arraystretch}{1.5}{
\resizebox{1\textwidth}{!}{%
\begin{tabular}{||c||c||c|c||c||c||}

\hline {\multirow{2}{*}{\textbf{Scheme} }}   & {\multirow{2}{*}{\textbf{Communication}}} & \multicolumn{2}{|c||}{\textbf{Computation}} &  \multicolumn{1}{|c||}{\multirow{2}{*}{\textbf{Setting}}} & {\multirow{2}{*}{\textbf{Privacy}}}\\ \cline{3-4}
 &  &  {$\boldsymbol  \db$} & {$ \boldsymbol \su$}& & \\ \hline

\hline \hline  \chorScheme &  $(\dbrow + \dbblock)\cdot \ns$ &  $\dbsize t_\oplus$ & $ (\dbrow+\dbblock)\cdot((\ns-1)\cdot t_\oplus)$ & \ns~\db s & $(\ns-1)${\em-private}\\

\hline  \goldbergScheme & $\dbrow\cdot\word\cdot\ns + \kr\cdot\dbblock$ & $(\dbsize/\word)\cdot t_\oplus$ & $\ns\cdot(\ns-1)\cdot\dbrow t_{\oplus} + 3\ns\cdot(\ns+1)t_{\oplus}$ & \ns~\db s& \tp{\em-private \ns-comp.-private} \\

\hline  \hline \PrSpec~\cite{gao2013location} & $(2\sqrt{\dbrow}+3)\cdot\lceil\log p\rceil$  & $\mathcal{O}(\dbrow)\cdot Mulp$ & $4\sqrt{\dbrow}\cdot Mulp$& $1$ \db & {\em underlying \pir~ broken}\\
\hline  Troja et al~\cite{troja2015efficient} & $12\delta\cdot\dbblock$ & $\mathcal{O}(\dbsize)\cdot Mulp$ & $4\sqrt{\dbsize}\cdot Mulp$ & $1$ \db & {\em computationally-private}\\ 

\hline  Troja et al~\cite{troja2014leveraging} & $n_g\cdot \pi\cdot\log_2q + (2\sqrt{\dbsize}+3)\cdot\lceil\log p\rceil$ & $\mathcal{O}(\dbsize)\cdot Mulp$ & $n_g\cdot \pi\cdot(2Expp+Mulp)+4\sqrt{\dbsize}\cdot Mulp$& $1$ \db & {\em computationally-private}\\

\hline  XPIR~\cite{aguilar2016xpir} & $ d\cdot(\dbrow/\alpha)^{1/d}\cdot \mathcal{C} + \lambda\cdot F^d\cdot\dbblock$ & $2d\cdot(\dbrow/\alpha)\cdot(\dbblock/\ell_0)\cdot Mulp$ & $d\cdot(\dbrow/\alpha)^{1/d}\cdot Enc + d\cdot\alpha\cdot\dbblock/\ell_0\cdot Dec$ & $1$ \db & {\em computationally-private}\\ \hline
\end{tabular}}}

\flushleft{\scriptsize{\textbf{Variables:} $t_\oplus$ is the execution time of one XOR operation. $p$ is a large prime, and $Mulp$ and $Expp$ are the execution time of performing one modular multiplication, and one modular exponentiation respectively. $\pi$~denotes the number of bits that an \su~shares with other \su s in~\cite{troja2014leveraging}, $n_g$ is the number of \su s within a same group in~\cite{troja2014leveraging}. $\delta$ is the number of \db~segments in~\cite{troja2015efficient}. $d$ is the recursion level, $\alpha$ is the aggregation level, $\mathcal{C}$ is the Ring-LWE ciphertext size, $\lambda$ is the number of elements returned by \db, $F$ is the expansion factor of the Ring-LWE cryptosystem, $\ell_0$ is the number of bits absorbed in a cyphertext, all are used in~\cite{aguilar2016xpir}. $(Enc,Dec)$ are respectively the encryption and decryption cost for Ring-LWE cryptosystem used in~\cite{aguilar2016xpir}.

}}
\end{table*}

\begin{figure}[h!]
\center
\includegraphics[width=0.5\textwidth]{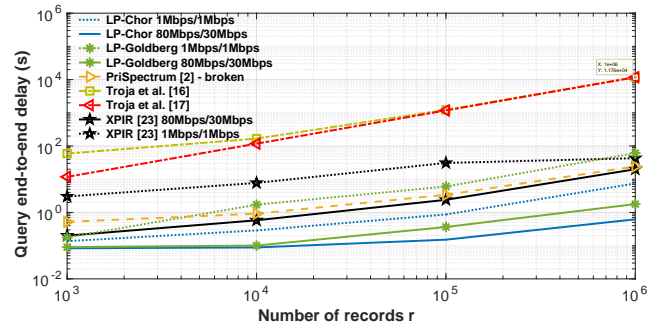} 
\caption{Query RTT of the different PIR-based approaches}
\label{fig:queryRTT}
\end{figure}

\noindent \textbf{Results and Comparison.} We first measure the query end-to-end delay of the proposed approaches and plot the results in \figurename~\ref{fig:queryRTT}. We also include the delay introduced by the existing schemes based on our estimation of the operations included in Table~\ref{tab:Table1}. The end-to-end delay that we measure takes into consideration the time needed by \su~to generate the query, the network delay, the time needed by \db~to process the query, and finally the time needed by \su~to extract the $\ind^{th}$ record of the database. We consider two different internet speed configurations in our experiment. We first rely on a high-speed internet connection of $80 Mbps$ on the download and $30 Mbps$ on the upload for all compared approaches. Then we use a low-speed internet connection of $1 Mbps$ on the upload and download to assess the impact of the bandwidth on \chorScheme~and \goldbergScheme, and also on XPIR as well.

\figurename~\ref{fig:queryRTT} shows that the proposed schemes perform much better than the existing approaches in terms of delay even with low-speed internet connection. They also perform better than the fastest existing \cpir~protocol XPIR. This shows the benefit of relying on multi-server \itpir~in multi-\db~\crn s. Also, and as expected, \chorScheme~scheme performs better than \goldbergScheme~thanks to its simplicity. As we will see later, \goldbergScheme~also incurs larger communication overhead than \chorScheme~as well. This could be acceptable knowing that \goldbergScheme~can handle collusion of up-to \ns~\db s, and is robust in the case of $(\ns-\kr)$ non-responding \db s, and \vbr~byzantine \db s, as opposed to \chorScheme. This means that \goldbergScheme~could be more suitable to real world scenario as failures and byzantine behaviors are common in reality. \figurename~\ref{fig:queryRTT} also shows that the network bandwidth has a significant impact on the end-to-end latency. This is due to the relatively large amount of data that needs to be exchanged during the execution of these protocols which requires higher internet speeds.

\begin{figure}[!h]
    \centering
    \subcaptionbox{\small SU Computational Overhead.\label{fig:suComuputation}}
    {\includegraphics[width=0.23\textwidth]{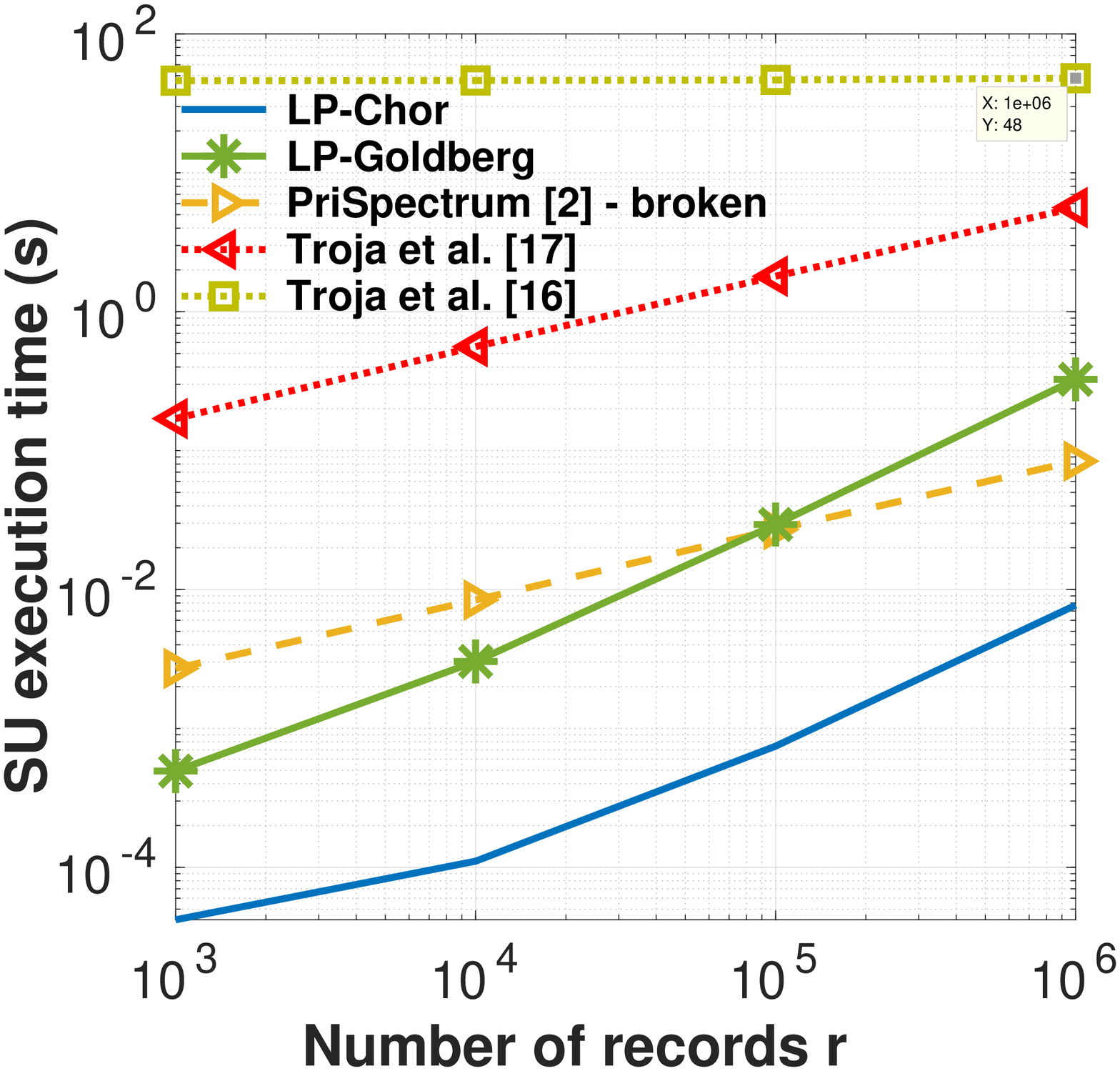}}\quad
    \subcaptionbox{\small DB Computational Overhead.\label{fig:dbComuputation}}
     {\includegraphics[width=0.23\textwidth]{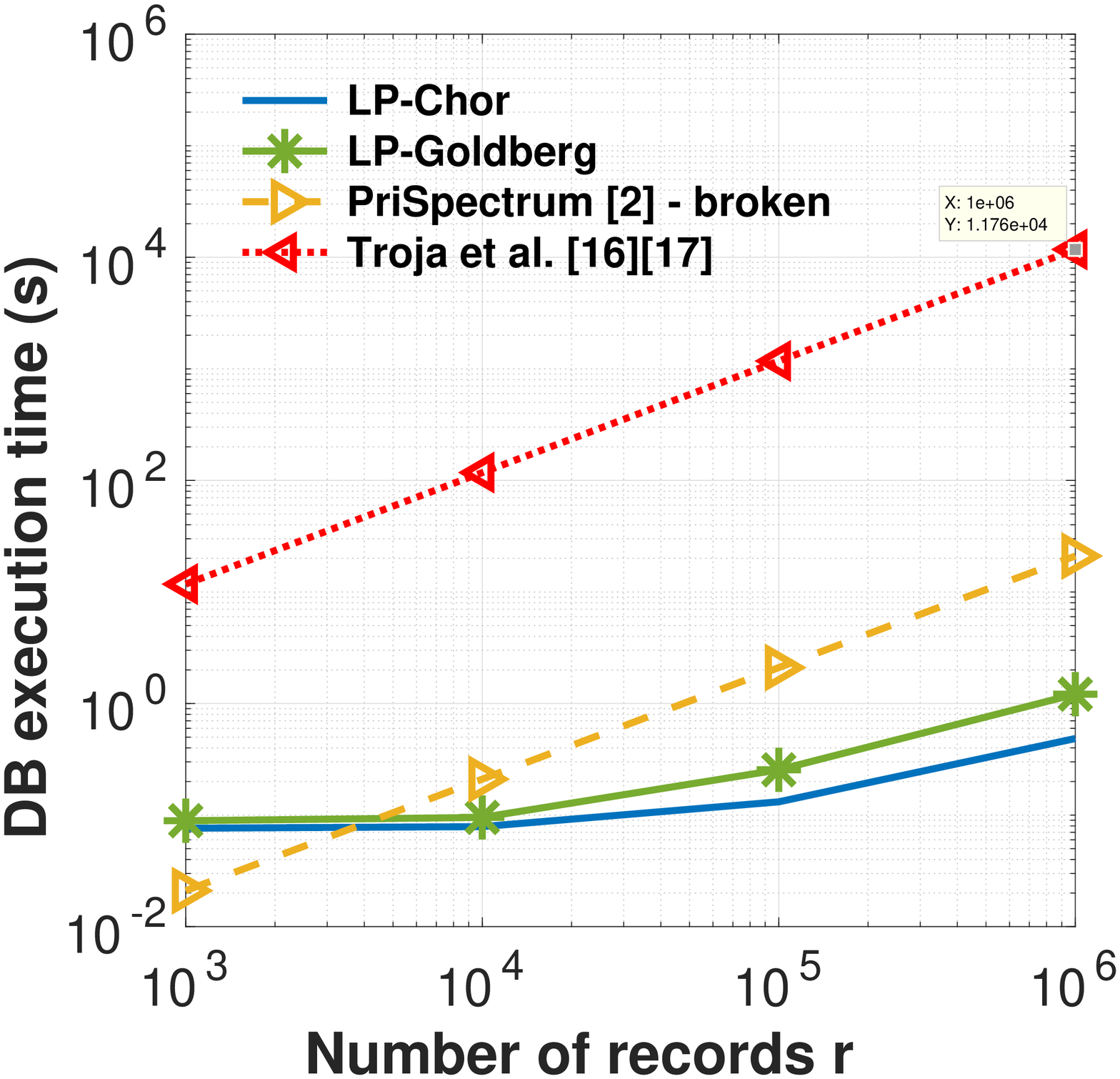}}
    \caption{Computation Comparison}
    \label{fig:comp}
\end{figure}

We also compare the computational complexity experienced by each \su~and \db~separately in the different approaches as shown in Table~\ref{tab:Table1}. We further illustrate this through experimentation and we plot the results in \figurename~\ref{fig:suComuputation}, which shows that the proposed schemes incur lower overhead on the \su~than the existing approaches. The same observation applies to the computation experienced by each \db~which again involves only efficient XOR operations in the proposed schemes. We illustrate this in \figurename~\ref{fig:dbComuputation}.

\begin{figure}[h!]
\center
\includegraphics[width=0.25\textwidth]{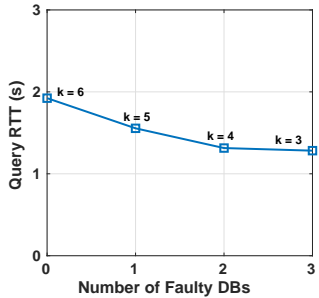} 
\caption{Impact of the number of faulty \db s on the query RTT.}
\label{fig:robustness}
\end{figure}

We also study the impact of non-responding \db s on the end-to-end delay experienced by the \su~in \goldbergScheme~as illustrated in \figurename~\ref{fig:robustness}. This Figure shows that as the number of faulty \db s increases, the end-to-end delay decreases since \su~needs to process fewer shares to recover the record $\dbmatrix_\ind$. As opposed to \chorScheme, in \goldbergScheme, \su~is still able to recover the record \ind~even if only \kr~out-of-\ns~\db s respond. Please recall also that our experiment was performed on resource constrained VMs to emulate \db s, however in reality, \db s should have much more powerful computational resources than of those of the used VMs which will have a tremendous impact on further reducing the overhead of the proposed approaches.

\begin{figure}[h!]
\center
\includegraphics[width=0.5\textwidth]{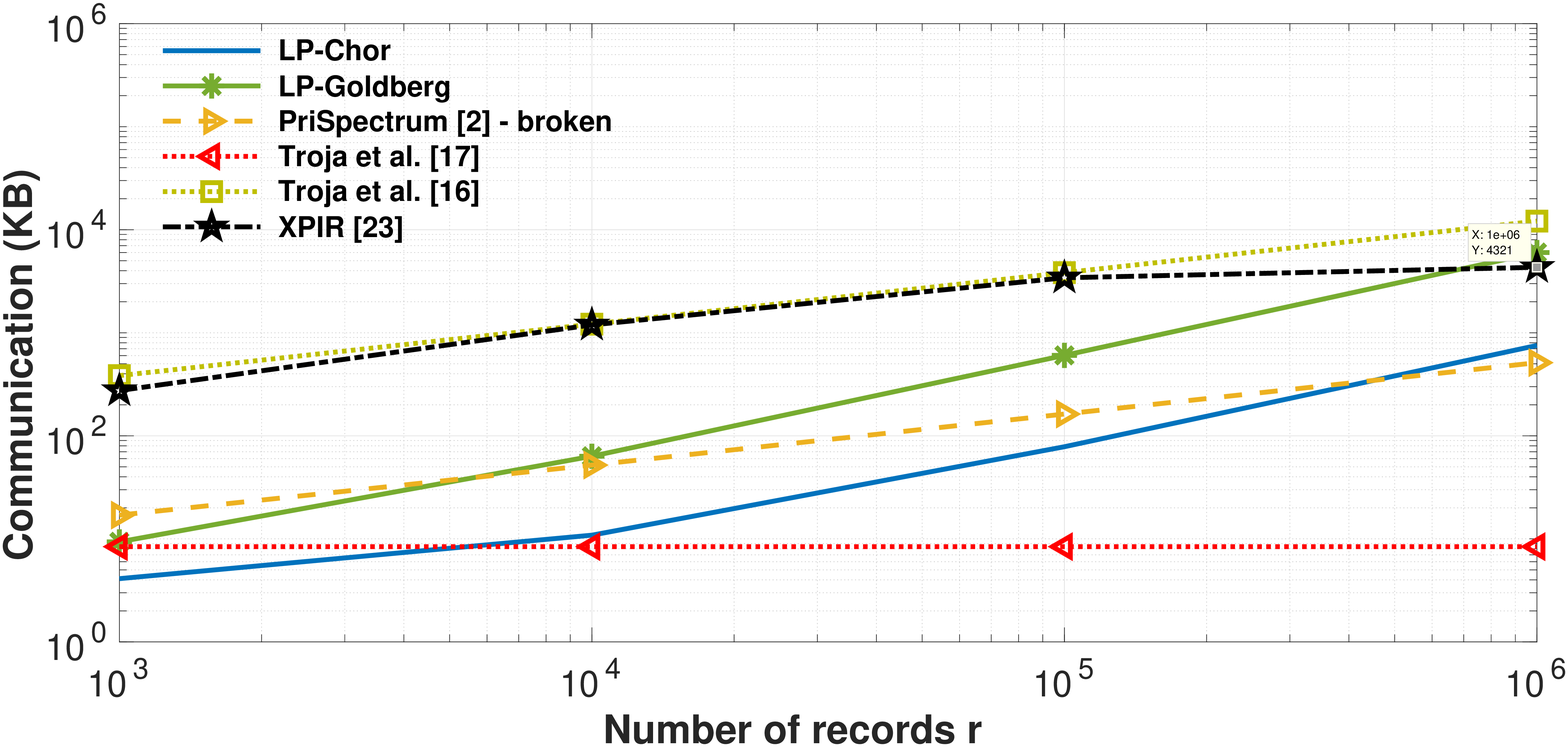}
 \caption{Comparison of the communication overhead of the different approaches: $\dbblock = 560$ B, $\kr = \ns$, $\vbr = 0$.}
\label{fig:comm}
\end{figure}

In terms of communication overhead, most of the approaches, including \chorScheme~and \goldbergScheme, have linear cost in the number of records in the database as shown in Table~\ref{tab:Table1}. What really makes a difference between these schemes' communication overheads is the associated constant factor which could be very large for some protocols. Based on our experiment and the expressions displayed in Table~\ref{tab:Table1}, we plot in \figurename~\ref{fig:comm}, the communication overhead that the \crn~experiences~for each private spectrum availability query issued by \su~for the different schemes. The scheme with the lowest communication overhead is that of Troja et al.~\cite{troja2015efficient} thanks to the use of Gentry et al. \pir~\cite{gentry2005single} which is the most communication efficient single-server protocol in the literature having a constant communication overhead. However this scheme is computationally expensive just like most of the existing \cpir-based approaches as we show in \figurename~\ref{fig:queryRTT}. \chorScheme~is the second best scheme in terms of communication overhead but it also provides information theoretic privacy. As shown in \figurename~\ref{fig:comm}, \chorScheme~incurs much lower communication overhead than \goldbergScheme~thanks to the simplicity of the underlying Chor~\pir~protocol. However, as we discussed earlier, \goldbergScheme~provides additional security features compared to \chorScheme. XPIR has a relatively high communication overhead especially for smaller database size but its overhead becomes comparable to that of \goldbergScheme~when the database's size gets larger as shown in \figurename~\ref{fig:comm}. This could be a good alternative to the \cpir~schemes used in the context of \crn s especially that it introduces much lower latency which is critical in the context of \crn s. Still, the proposed approaches have better performance and also provide information-theoretic privacy to \su s, which shows their practicality in real world. 

\section{Related Work}
\label{sec:Related}
There are other approaches that address the location privacy issue in database-driven \crn s. However, for the below mentioned reasons we decided not to consider them in our performance analysis.  For instance, Zhang et al.~\cite{zhang2015optimal} rely on the concept of {\em k-anonymity} to make each \su~queries \db~by sending a square cloak region that includes its actual location. {\em k-anonymity} guarantees that \su's location is indistinguishable among a set of $k$ points. This could be achieved through the use of dummy locations by generating $k-1$ properly selected dummy points, and performing $k$ queries to \db, using the real and dummy locations. Their approach relies on a tradeoff between providing high location privacy level and maximizing some utility. This makes it suffer from the fact that achieving a high location privacy level results in a decrease in spectrum utility. However, {\em k-anonymity}-based approaches cannot achieve high location privacy without incurring substantial communication/computation overhead. Furthermore, it has been shown in a recent study led by Sprint and Technicolor~\cite{zang2011anonymization} that anonymization based techniques are not efficient in providing location privacy guarantees, and may even leak some location information. Grissa et al~\cite{grissa2015cuckoo} propose an information theoretic approach which could be considered as a variant of the trivial \pir~solution. They achieve this by using set-membership probabilistic data structures/filters to compress the content of the database and send it to \su~which then needs to try several combinations of channels and transmission parameters to check their existence in the data structure. However, LPDB is only suitable for situations where the structure of the database is known to \su s which is not always realistic. Also, LPDB relies on probabilistic data structures which makes it prone to false positives that can lead to erroneous spectrum availability decision and cause interference to \pu's transmission. 
Zhang et al.~\cite{zhang2015achieving} rely on the {\em $\epsilon$-geo-indistinguishability} mechanism~\cite{andres2013geo}, derived from {\em differential privacy} to protect bilateral location privacy of both \pu s and \su s, which is different from what we try to achieve in this paper. This mechanism helps \su s obfuscate their location, however, it introduces noise to \su's location which may impact the accuracy of the spectrum availability information retrieved.


\section{Conclusion}
\label{sec:Conclusion}
In this paper, with the key observation that database-driven \crn s contain multiple synchronized \db s having the same content, we harnessed multi-server \pir~techniques to achieve an optimal location privacy for \su s with high efficiency. Our analytical and experimental analysis indicate that our adaptation of multi-server \pir~for database-driven \crn s achieve magnitudes of time faster end-to-end delay compared to the fastest state-of-the-art single-server \pir~adaptation with an information theoretical privacy guarantee. Specifically, we adapted two multi-server \pir~techniques into \crn~settings as \chorScheme~and \goldbergScheme. \chorScheme~achieves an end-to-end delay below a second with high collusion resiliency, while \goldbergScheme~offers fault tolerance and byzantine robustness with a significantly higher efficiency compared to single-server \pir~based approaches. Given the demonstrated benefits of multi-server \pir~approaches without incurring any extra architectural overhead on database-driven \crn s, we hope this work will provide an incentive for the research community to consider this direction when designing location privacy preservation protocols for \crn s. 

\small{
\bibliographystyle{IEEEtran}
\bibliography{./references,references-database,references-database-consistency,refs-security-privacy-sensing-14-15,refs-bechir-privacy-wireless-systems,references-TCCN-16-Mohamed}
}

\end{document}